\documentstyle[preprint,aps,prb]{revtex}

\begin{document}
\draft
\preprint{}
\title{A Computational Study of Thirteen-atom Ar-Kr Cluster \\
    Heat Capacities}
\author{D. D. Frantz}
\address{Department of Chemistry, University of Waterloo, \\
    Waterloo, Ontario N2L 3G1, Canada}
\date{\today}
\maketitle
\begin{abstract}
Heat capacity curves as functions of temperature were calculated using Monte
Carlo methods for the series of Ar$_{13-n}$Kr$_n$ clusters ($0 \leq n \leq
13$). The clusters were modeled classically using pairwise additive
Lennard-Jones potentials. J-walking (or jump-walking) was used to overcome
convergence difficulties due to quasiergodicity present in the solid-liquid
transition regions, as well as in the very low temperature regions where heat
capacity anomalies arising from permutational isomers were observed.
Substantial discrepancies between the J-walking results and the results
obtained using standard Metropolis Monte Carlo methods were found. Results
obtained using the atom-exchange method, another Monte Carlo variant designed
for multi-component systems, were mostly similar to the J-walker results.
Quench studies were also done to investigate the clusters' potential energy
surfaces; in each case, the lowest energy isomer had an icosahedral-like
symmetry typical of homogeneous thirteen-atom rare gas clusters, with an Ar
atom being the central atom.
\end{abstract} \pacs{}
\narrowtext

\section{Introduction}
The study of clusters has shown these systems to exhibit a surprisingly rich
diversity of interesting phenomena. It has helped provide much insight into
the transition from finite to bulk behavior, since many of the physical
properties of bulk systems are highly modified in clusters because of their
much greater fraction of surface atoms. Although most studies have dealt with
homogeneous clusters, heterogeneous clusters have been increasingly the
subjects of both
experimental\cite{Pd-Pt,HPGB,KHTYHNK,SZGKZ,BHBEJ,Knickelbein,ZLW,WLTD} and
theoretical\cite{LMA,FF,GPC,GHJ,TZKDS,TAP,Lopez-Freeman,CSA,CKP,HBBEJ,LGNS,RA,AS,GLKW,LeRoy}
investigations. Most of these studies have concentrated on bimetallic
clusters, in large part because of the roles they play in catalysis.
Heterogeneous clusters are also important theoretically, since the cluster
composition provides another variable that can be exploited to better
understand the transition from microscopic to macroscopic behavior.
Much of the intriguing behavior of homogeneous clusters, such as their varied
melting
behavior\cite{coexist,DJB,J-walker,Tsai-Jordan,HA,EK,BB,BBDJ,Wales,HMD,AS2,GBB,BJ,VSA}
and the appearance of ``magic number'' effects in many cluster properties as
functions of aggregate cluster size,\cite{Northby,FD_magic,ESR,Magic_Cv}
is also observed in heterogeneous clusters, but this behavior can be strongly
dependent on the cluster composition.

It is the desire to better understand the role cluster composition plays in
the thermodynamic properties of heterogeneous clusters that has motivated my
present work, and so I have used binary rare gas clusters as my subject.
These systems can be adequately described by a simple pairwise additive
Lennard-Jones potential, unlike bimetallic clusters, which require more
complicated (and computationally expensive) potentials for proper
treatment.\cite{LMA} I have also chosen Ar and Kr as the two components
because their atomic sizes are similar enough that most of the physical
properties of the homogeneous clusters are not drastically altered in going
to binary clusters of similar aggregate size; a related study of Ne-Ar
clusters, where the atomic sizes are sufficiently dissimilar that the
physical behavior of the binary clusters differs radically from that of their
homogeneous counterparts, is nearing completion and will be submitted for
publication shortly.

I have also limited the current study to thirteen-atom clusters.
Thirteen-atom
homogeneous\cite{LeRoy,coexist,DJB,J-walker,Tsai-Jordan,HA,EK,BB,Magic_Cv,NP,QS}
and heterogeneous\cite{LMA,FF,GPC,GHJ,TZKDS,TAP,Lopez-Freeman} clusters have
been the subject of many studies. The homogeneous clusters exhibit ``magic
number'' effects for many of their properties that are a manifestation of
their compact icosahedral ground state configuration. Because Ar and Kr have
similar sizes (the Ar radius is about 11\% smaller than the Kr radius),
thirteen-atom Ar-Kr clusters also have icosahedral-like lowest-energy
configurations that are only slightly distorted; Tsai, Abraham and
Pound\cite{TAP} found the lowest-energy configuration of Ar$_7$Kr$_6$ to be
icosahedral-like, and Lopez and Freeman\cite{Lopez-Freeman} found the
ground state configuration of Pd$_6$Ni$_7$ clusters to be icosahedral-like
as well (the Ni-Pd radius ratio is very similar to the Ar-Kr radius ratio).
This size similarity then leads to two different categories of isomers:
topological isomers based on geometric structures that are similar to those
of their homogeneous counterparts, and permutational isomers, which are based
on the various rearrangements of the different component atoms within a
topological form. While homogeneous clusters have only topological isomers,
heterogeneous clusters can have several permutational isomers associated with
each topological form, depending on the cluster composition. For
Pd$_6$Ni$_7$, Lopez and Freeman found that neither changing the potential
model nor the Pd-Ni interaction strength had any effect on the topological
form of the ground state isomer (which was always icosahedral-like), but did
strongly effect which of the permutational isomers had the lowest energy,
with the segregated isomer being lowest for lower Pd-Ni interaction
strengths, and the mixed isomer being lowest for higher interaction
strengths. Similar results were obtained by L\'{o}pez, Marcos and
Alonso\cite{LMA} in their study of thirteen and fourteen-atom Cu-Au clusters.

One of the more sensitive cluster properties is the heat capacity, which has
been useful in elucidating the nature of cluster solid-liquid ``phase''
transitions.\cite{Lopez-Freeman,DJB,J-walker,Tsai-Jordan,Magic_Cv,NP,QS,Matro_Freeman,Lopez,JWFPI}
Heat capacity curves as functions of temperature for homogeneous
thirteen-atom clusters are characterized by a very large peak in the
solid-liquid transition region that is a consequence of the large energy gap
between the ground state icosahedral isomer and the higher energy
non-icosahedral isomers.\cite{DJB,J-walker,Tsai-Jordan,Magic_Cv,NP,QS} Lopez and Freeman\cite{Lopez-Freeman}
showed that the heat capacity curve for Pd$_6$Ni$_7$ is quite similar to the
homogeneous case, which is largely due to the two having similar topological
isomers. However, they also observed the effects of the permutational
isomers, made manifest as an additional, very small, heat capacity peak
occurring at very low temperatures. This feature, arising from the low-energy
icosahedral-like isomers is reminiscent of order-disorder transitions known
to occur in some bulk alloy materials.

Cluster heat capacities are notoriously difficult to calculate accurately
using simulation methods because of poor convergence. This is due to
quasiergodicity, or the incomplete sampling of configuration space, which
arises in systems characterized by well separated regions that are linked by
very low transition probabilities because of bottlenecks in the configuration
space. Thus the sampling is effectively confined to only a small subset of
the important regions.\cite{Val-Whit} In the past, this has necessitated the
use of extremely long simulations to ensure proper sampling, and it has only
been recently that advances in computer technology have made accurate heat
capacity calculations from simulations feasible. In addition, new methods
have been developed that substantially reduce quasiergodicity and
dramatically increase convergence in Monte Carlo simulations. These methods
are characterized by their use of large scale movements, or jumps, of the
random walker to various regions of configuration space and so are denoted
``J-walking'' or ``jump-walking.''\cite{J-walker} J-walking has been
used successfully in several
studies,\cite{Tsai-Jordan,Magic_Cv,Matro_Freeman,Lopez,JWFPI,SLF} including
the Lopez and Freeman work,\cite{Lopez-Freeman} where it not only worked well
in the Pd$_6$Ni$_7$ solid-liquid transition region, but also in the very low
temperature mixing-anomaly region; the mixing-anomaly peak was not obtained
using the usual Metropolis method,\cite{MRT2} but was seen only when
J-walking was used. J-walking was used in this study as well.

I begin in Section~\ref{Sec:theory} with a description of the various
computational methods I used, including a brief discussion of my
implementation of the J-walking method for binary clusters.
Section~\ref{Sec:results} first examines the structural properties of the
thirteen-atom Ar-Kr clusters, then discusses the results obtained using
J-walking for the heat capacities and potential energies as functions of
temperature. These are compared to the results obtained from similar
calculations done using the standard Metropolis method, and to the results
obtained from a modified Metropolis scheme incorporating an atom-exchange
algorithm.\cite{TAP} While atom-exchange techniques provided substantial
improvement over the unaided Metropolis method for the heat capacities at
very low temperatures, they were still unable to overcome the quasiergodicity
difficulties nearly as well as the J-walking method. The results of quench
studies are also discussed. Finally, in Section~\ref{Sec:conclusion}, I
summarize my findings and discuss some of the insights provided by the binary
cluster results.

\section{Computational Method}                        \label{Sec:theory}
Monte Carlo simulations were run for all clusters in the series
Ar$_{13-n}$Kr$_n$ with $0 \leq n \leq 13$. The clusters were modeled by the
usual pairwise additive Lennard-Jones potential,
\begin{eqnarray}
    V & = & \sum_{i < j} V_{LJ}(r_{ij}), \nonumber \\
    V_{LJ}(r_{ij}) & = & 4\epsilon_{ij}\left[\left(
        \frac{\sigma_{ij}}{r_{ij}}\right)^{12}
        - \left(\frac{\sigma_{ij}}{r_{ij}}\right)^6 \right].
\end{eqnarray}
Table~\ref{Tbl:LJ-params} lists the Lennard-Jones parameters.\cite{LDW} The
Ar-Kr interaction parameters were determined from the usual Lorentz-Berthelot
mixing rules;\cite{Allen-Tild} the reader is referred to the work by Lopez
and Freeman concerning the effects of using generalized mixing rules for
Pd$_6$Ni$_7$.\cite{Lopez-Freeman} The classical internal energy and heat
capacity were calculated by the usual expressions for an $N$-atom cluster,
\begin{eqnarray}
    \left\langle U^* \right\rangle & = & \frac{3NT^*}{2}
        + \left\langle V^* \right\rangle, \\
    \left\langle C_V^* \right\rangle & = & \frac{3N}{2}
    + \frac{\left\langle{(V^*)^2}\right\rangle
    - \left\langle{V^*}\right\rangle^2}{(T^*)^2}.
\end{eqnarray}
The reduced units are in terms of the Ar-Ar interaction, with $U^* =
U/\epsilon_{\mbox{\scriptsize Ar-Ar}}$, $V^* = V/\epsilon_{\mbox{\scriptsize
Ar-Ar}}$, $C_V^* = C_V/k_B$, and $T^* = k_BT/\epsilon_{\mbox{\scriptsize
Ar-Ar}}$.

Because of their finite vapor pressure, small clusters are known
to become unstable at sufficiently high temperatures.\cite{QS,EK1} For
clusters modeled with the Lennard-Jones potential under free volume
conditions, the average energy vanishes in the limit of infinitely long
walks. Consequently, the choice of boundary conditions can greatly effect
some cluster properties at higher temperatures.\cite{BBDJ} I have followed
Lee, Barker and Abraham\cite{LBA} and have confined the clusters by a
perfectly reflecting constraining potential of radius $R_c$ centered on the
cluster's center of mass. To maintain a common set of boundary conditions
throughout the survey, the constraining radius was identical for all the
clusters studied, with $R_c = 3\sigma_{\rm Kr-Kr}$. As will be described
later, the use of a constraining sphere for the simulation of binary clusters
led to some computational difficulties that required special handling.

\subsection{J-Walking}
Standard Monte Carlo simulations of clusters based on the sampling algorithm
proposed by Metropolis {\em et al.}\cite{MRT2} are known to suffer from
systematic errors due to quasiergodicity,\cite{Val-Whit} the non-ergodic
sampling that typically arises with configuration spaces that are comprised
of several regions separated by large barriers. For certain temperature
domains, this can lead to bottlenecks that effectively confine the sampling
to only some of the regions, resulting in large errors.\cite{Allen-Tild}
These systematic errors are purely a consequence of the practical
limitations placed on the walk lengths, since, for sufficiently long walks,
the random walker overcomes the bottlenecks often enough to provide ergodic
sampling. The required walk length depends on the temperature; for higher
temperatures, the random walker's stepsizes are large enough that movement
between the various regions occurs frequently and so shorter walk lengths are
adequate. As the temperature decreases, so does the probability of moving
between regions, thus requiring that the walk lengths be increased. The
potential hypersurfaces for clusters typically have deep wells corresponding
to very stable, compact isomers that are separated from one another by large
barriers. For high temperatures corresponding to the cluster's ``liquid''
region where unhindered isomerization occurs, the random walker has full
access to all of configuration space. However, for temperatures in the
solid-liquid transition region, a dichotomy of time scales characterizes
the random walks, producing rapid movement within the isomeric regions, but
only very slow movement between them.

J-walking addresses the problem of quasiergodicity by coupling the
usual small-scale Metropolis moves to occasional large-scale jumps
that move the random walker to the different regions of configuration space
in a representative manner.\cite{J-walker} In the following paragraphs, I
give a brief summary of the method and describe some of the aspects that were
important in implementing the method for heterogeneous clusters; the reader
is referred to the original papers for a more complete description of the
general algorithms.\cite{J-walker,JWFPI}

The large-scale jumps are governed by Boltzmann distributions generated at
higher temperatures where the sampling is ergodic. There are two
complementary implementations for generating the classical J-walker Boltzmann
distributions. The first runs the high temperature walker (J-walker) in
tandem with the low temperature walker, with the low temperature walker
occasionally attempting jumps to the current J-walker position simply by
using the current J-walker coordinates as its trial position. The high
computational cost of this implementation limits its practical use to
parallel computers. In the second implementation, the J-walker is run
beforehand and the configurations generated during the walk are stored
periodically in an external array.  Subsequent jump attempts are made by
accessing the stored configurations via randomly generated indices. This
implementation has only a modest computational overhead (mostly the time
required to generate the distributions), but requires very large storage
facilities for handling the distribution arrays.  Because fast workstations
(and even personal computers now) having tens of Mb of RAM and Gb of disk
storage are affordable and quite common, while access to parallel computers
is still much more limited, this implementation remains the method of choice
and was the one used for all the J-walking calculations reported in this
study.\cite{computers} The reader is referred to a recent study of ammonium
chloride clusters by Matro, Freeman and Topper for a description of a
parallel J-walking implementation that combined the tandem-walker and
external-file methods.\cite{Matro_Freeman}

For physically realistic systems such as clusters, the Boltzmann
distributions are generally too narrow for a single distribution to span the
entire temperature domain, and so the distributions have to be generated in
stages. For each cluster studied, an initial J-walker distribution was
generated from a long Metropolis walk at a temperature high enough for the
sampling to be ergodic. This distribution was then used for J-walking runs to
obtain averages of the potential energy and the heat capacity for a series of
lower temperatures. The J-walker distribution's width placed practical
limitations on the effective temperature range for the subsequent J-walking
simulations. When the temperature difference between the distribution and the
low temperature walker became too large, very few attempted jumps were
accepted since the J-walker configurations likely to be accepted were in
the low energy tail of the distribution. Thus, at the temperature where the
jump acceptance became too small,\cite{acceptance} a new distribution was
generated from the previous one, using J-walking to ensure it was ergodic as
well. This distribution was then used to obtain data for the next lower
temperature range, and then to generate the next lower temperature J-walker
distribution, and so on, until the entire temperature domain was spanned.
The distributions used in this study each consisted of at least $10^6$
configurations. Because the configurations generated during Metropolis
walks are highly correlated, they were stored in a periodic fashion.
The higher temperature distributions were the widest, and contained the
greatest variety of configurations; these were generated by storing
configurations every hundred passes. For the narrower lower temperature
distributions, the sampling was reduced to every fifty or twenty-five passes,
and for the very narrow distributions for the low temperatures corresponding
to the solid region, a sampling of every ten passes was sufficient.

The J-walker calculations were done in a manner mostly analogous to a
previous J-walker study I did on a series of homogeneous clusters
ranging in size from four to twenty-four atoms.\cite{Magic_Cv} In that study,
a larger constraining radius ($R_c = 4\sigma$) was chosen to freely
accommodate the larger clusters, and the initial J-walker distributions were
all generated at higher temperatures corresponding to the cluster
dissociation region. This was done to eliminate small effects of
quasiergodicity occurring in the liquid-dissociation transition region for
some of the larger clusters.  Preliminary calculations indicated that using
such high initial temperatures for small binary clusters resulted in problems
with the cluster constraining potential due to the simulations not using
center of mass coordinates. For such simulations, the center of mass (and
thus the cluster as a whole) will drift during the course of the walk,
requiring that the center of mass coordinates be updated for each accepted
move. This can be easily done, and the computational overhead is quite low.
In this scheme, attempted Metropolis moves are checked for constraining
potential violations by calculating the new center of mass consistent with
the trial move and checking if the distance to the new center of mass is
greater than the constraining radius $R_c$. If it is, the attempted move
is rejected outright, otherwise, the usual Metropolis criterion is used to
determine whether the move is accepted or not. While such a scheme guarantees
that the trial atom is never moved outside the constraining potential, it
does not guarantee that an accepted Metropolis move does not leave one of the
other cluster atoms outside the constraining potential when the move is
accepted and the center of mass updated.  Such an orphaned atom can then
drift away from the rest of the cluster, effecting the results for the
remainder of the Metropolis simulation.  Although this is a rare occurrence,
it can become a significant problem at very high temperatures where the
cluster is completely dissociated (thus physically corresponding to a highly
compressed gas confined to a small spherical cavity) and there is greater
likelihood of atoms being near the constraining surface. The problem is also
worse for smaller clusters since these can have larger changes in the center
of mass when moves are accepted. Heterogeneous cluster simulations are
especially prone to such constraining sphere violations since the
interactions between the different components can be quite disparate, leading
to the dissociation of one component at temperatures where the other
components remain intact.

The constraining potential violations are not a problem for J-walking
simulations, provided that the J-walker distributions are free of such
violating configurations. This is because a J-walker ``walk'' is really
a series of short ``excursions,'' each starting from the J-walker
configuration last accepted. Thus, should a Metropolis move ever result
in an atom being left outside the constraining potential, the next
accepted jump would restore the configuration to a valid one. This does
require that much care be exercised in ensuring that proper J-walker
distributions have been generated from the initial high temperature
Metropolis walks.  So, for each cluster studied, the initial J-walker
distributions were generated at temperatures in the cluster liquid region
(from 44 K for Ar$_{12}$Kr to 58 K for ArKr$_{12}$), well below the cluster
dissociation region. Even at these temperatures, single atom dissociations
occurred frequently enough that a few constraining potential violations
occurred. Sampled configurations for these high temperatures were stored in
the J-walker distribution files only every hundred passes, and since checking
every atom for possible constraining potential violations for every accepted
move is computationally expensive, and considering that the frequency of
violations was very low, I instead found it preferable to simply generate
extra configurations and then check every configuration in the distribution,
discarding any violating configurations (typically less than 0.1\% of the
configurations stored).

Correlations in the J-walker distributions were further reduced by writing
the distribution files in a parallel fashion. All output distribution files
were opened at the start of the program and configurations were written to
each in turn, rather than writing to each file sequentially, one at a time.
Thus, each distribution file contained configurations sampled from the entire
run. Although this resulted in highly fragmented files, which could have
greatly increased the time required to later read a distribution into memory,
this did not turn out to be a problem because as each distribution file was
subsequently checked for constraining potential violations, the valid
configurations were written to the disk sequentially.

Preliminary calculations also indicated another important difference
between the binary cluster simulations and the homogeneous cluster
calculations done previously.\cite{Magic_Cv} In those calculations,
comparison of the J-walking heat capacity results for a given cluster
size that were obtained from different, independently generated
distributions were in agreement with one another, implying that the
distributions were sufficiently representative and thus any systematic
errors were negligible. Those results were obtained from averages collected
from $10^7$ passes for each temperature. For the binary clusters however,
there were small differences between the J-walking heat capacity results
obtained from different, independently generated sets of distributions,
indicating that small systematic errors were present in the J-walking heat
capacity data that were larger than the uncertainties associated with random
fluctuations obtained with $10^7$ passes. This was most likely a consequence
of the large increase in the number of isomers in the binary cluster case
because of the extra permutational isomers, which made obtaining
representative samples that much more difficult. To ensure there were no
significant systematic errors in the binary cluster heat capacity curves, I
therefore ran five separate J-walking trials, with each trial sampling from
its own unique set of J-walker distributions. The results for the five
separate trials were then combined and averaged, with the standard deviation
taken as an estimate of the uncertainty. For each trial, the total walk
length was set to $10^6$ passes of data accumulation, so that the total
computational time was not much longer than that for a single run of $10^7$
passes. Fig.~\ref{Fig:five_trials} shows the results for Ar$_8$Kr$_5$.
As can be seen in the plot, the noise level in each curve is about the same
magnitude as the differences between the curves, indicating that these walk
lengths were sufficiently long for the desired level of accuracy. Also shown
are the results obtained for two similar J-walking trials, each sampling from
a set of distributions initially generated at $T = 70$ K, a temperature
corresponding to the high temperature side of the cluster dissociation peak.
Systematic errors are clearly evident in these curves. Not only do they
differ substantially from the five J-walking curves generated from the
initially lower temperature J-walker distributions (especially at the very
low temperatures in the mixing-anomaly region), they do not even agree
between themselves, showing large differences in the solid-liquid transition
region.

\subsection{Standard Metropolis}
Since J-walking is a relatively new method, and experience with it is still
being obtained, I also ran standard Metropolis simulations for each cluster.
These provided a check of the J-walking results for those temperature regions
where quasiergodicity in the Metropolis runs was not a problem, as well as
revealing trends in the systematic errors arising from quasiergodicity in the
Metropolis runs as functions of cluster composition. Temperature scans were
generated with a mesh size of $\Delta T = 0.1$ or 0.2~K for very low
temperatures, and $\Delta T = 0.5$~K for higher temperatures. For each
temperature, simulations consisted of $10^5$ warmup passes, followed by
$10^7$ passes with data accumulation. The scans were started at the lowest
temperature using the global minimum configuration and were continued past
the cluster melting region, with the final configuration for each temperature
used as the initial configuration for each subsequent temperature. For almost
all of the clusters examined, the Metropolis and J-walking heat capacity
results agreed qualitatively throughout, except for very low temperatures
where small peaks associated with mixing anomalies were absent in the
Metropolis curves. They agreed quantitatively over much of the temperature
ranges as well, with the largest discrepancies occurring mostly in the
solid-liquid transition region effecting the peak height and location.
Interestingly, the agreement was worse for those clusters having a higher
proportion of Kr. Fig.~\ref{Fig:five_trials} also shows the Metropolis
results for Ar$_8$Kr$_5$. As can be seen, substantial discrepancies due to
quasiergodicity are evident in the solid-liquid transition region, and the
low-temperature mixing-anomaly peak is completely absent. Lopez and
Freeman\cite{Lopez-Freeman} obtained similar results in their study of
Pd$_6$Ni$_7$ clusters --- J-walking results showed a small mixing-anomaly
peak occurring at low temperatures that was absent in the Metropolis results.
The open circles in the plot represent Metropolis results obtained using
$10^8$ total passes. These are in agreement with the J-walker results
obtained from distributions initially generated at $T = 52$~K, verifying that
the J-walker distributions were free of systematic errors, as well as showing
that these Metropolis walks were sufficiently long to overcome
quasiergodicity in the solid-liquid transition region.

\subsection{Atom Exchange Method}
The inadequacy of the standard Metropolis algorithm for simulating
heterogeneous clusters at very low temperatures has been noted previously.
Tsai, Abraham, and Pound\cite{TAP} developed a simple but effective strategy
to help overcome these limitations by incorporating an atom exchange scheme
in their Metropolis algorithm. The method enhances the mixing of clusters by
occasionally attempting an exchange move where one of the atoms of one
component is swapped with one of the atoms of another component. The
move is either accepted or rejected according to the usual Boltzmann-weighted
criterion. To help resolve the discrepancies occurring between the standard
Metropolis and J-walking methods, I ran atom-exchange simulations for all
temperature regions where significant discrepancies were found. These were
done in a manner similar to the standard Metropolis runs described
previously, with an atom-exchange move between randomly selected Ar and Kr
atoms attempted once after every pass of standard Metropolis moves. The
results for Ar$_8$Kr$_5$ can also be seen in Fig.~\ref{Fig:five_trials}. The
atom-exchange results are a substantial improvement over the standard
Metropolis results in the low temperature region, but are only in qualitative
agreement with the J-walker results, showing a mixing-anomaly peak that is
much smaller than the J-walker peak and occurring at a higher temperature.
Nonetheless, the atom-exchange results lend further validity to the J-walker
results and demonstrate that the mixing-anomaly peak is not a spurious
artifact of J-walking.

\subsection{Quench studies}
In their study of the general cluster morphology of binary clusters, Clark
{\em et al.}\cite{CKP} conveniently characterized the general Lennard-Jones
potential parameter space for components A and B in terms of $\alpha =
\epsilon_{\rm AB}/\epsilon_{\rm AA}$, $\beta = \epsilon_{\rm
BB}/\epsilon_{\rm AA}$, $\Gamma = \sigma_{\rm AB}/\sigma_{\rm AA}$, and
$\Delta = \sigma_{\rm BB}/\sigma_{\rm AA}$. These parameters express the
relative interaction strengths and particle sizes for the cluster. Using the
Lorentz-Berthelot mixing rules reduces the parameter space dimensionality,
with $\alpha = \sqrt{\beta}$ and $\Gamma = \frac{1}{2}(1+\Delta)$, thus
restricting the cluster morphology to regions in the cluster ``phase''
diagram corresponding to roughly spherical shapes with varying degrees of
intermixing between the components, rather than elongated shapes with the two
components mostly segregated at either end. Using the values listed for Ar
and Kr in Table~\ref{Tbl:LJ-params}, these parameters become $\alpha =
0.8533$, $\beta = 0.7280$, $\Gamma = 0.9449$ and $\Delta = 0.8897$. For
medium sized clusters (about 50 to 250 particles) having equal numbers of Ar
and Kr atoms at temperatures in the liquid regime, these values of $\alpha$
and $\beta$ would correspond to systems mostly consisting of a Kr core
surrounded by Ar atoms if the atomic sizes were equal ($\Gamma = \Delta =
1$), but the values of $\Gamma$ and $\Delta$ would correspond to a system
mostly consisting of an Ar core coated by Kr atoms if the interatomic
interactions were equal ($\alpha = \beta = 1$). Clusters in the liquid regime
are sufficiently distended that small size differences between the atoms have
only a minor effect, but at the lower temperatures corresponding to the
cluster solid region, even small size differences can have a large influence
on cluster structure as packing effects become important. Thus there are
competitions between the intermolecular forces and atomic sizes that
determine the structural properties of binary clusters at various
temperatures, and ultimately their dynamical properties. For example, in
their study of 55-atom Ar-Kr clusters, Tsai, Abraham and Pound\cite{TAP}
described the competition between the Kr size that was a driving force for
its segregation to the surface, and its stronger bond strength that tended to
keep it in the interior. For both Ar$_{36}$Kr$_{19}$ and Ar$_{19}$Kr$_{36}$,
the Kr atoms tended to reside in the inner shells, while the Ar atoms tended
to segregate to the surface, although the central atom in each case was an Ar
atom.

At the lower temperatures associated with the cluster solid-liquid transition
regions, the competition between the intermolecular forces and atomic sizes
effects which of the multitude of isomeric structures will play dominant
roles in the cluster's behavior. Thus, information concerning the relative
populations of the lower energy isomers encountered during a simulation at a
given temperature can offer much insight into the nature of cluster
dynamics.\cite{SW,Amar_Berry} The relative frequencies of occurrence for each
isomer can be obtained from quench studies done during the simulation, where
the assumption is that the distribution of minima found in the quenches
indicates the likelihood of the system being associated with a particular
isomer. That is, the distributions are a reflection of the relative phase
volume of each catchment basin.\cite{WB} To obtain further insight into the
role cluster structure plays, I also performed quench studies on each of the
clusters simulated. Steepest-descent quenches\cite{SW} were performed
periodically on each cluster during one of its five J-walker trials to
monitor the relative frequency of occurrence for the lower energy isomers as
functions of temperature. For each temperature, quenches were undertaken
every 1000 passes, providing 1000 quenched configurations for subsequent
analysis. In each case, quench trajectories were run until the relative
energy difference converged to within $10^{-7}$.

Having a complete listing of all the stable isomers for a given cluster would
also be a valuable source of information. Unfortunately, the local minima
comprising a typical cluster potential hypersurface are far too numerous to
be completely catalogued in any practical manner. However, a reasonably
complete listing can be easily obtained from the J-walker distributions.
Since the distributions generated at a given temperature contain
representative samples of a cluster's configuration space, a crude, but
nonetheless effective, way to identify cluster isomers is to simply quench a
sufficient number of the configurations stored in the distributions, saving
them in an external file indexed by their energy, and then removing the
duplicate configurations. This was done for all the clusters examined in this
study. In each case, the configurations in one of the J-walker distribution
files generated at the initial J-walker temperature were quenched and
analyzed, and all the unique configurations saved to a file. Then the
configurations in the next lower temperature distribution were quenched, and
the unique configurations added to the file. This was continued until no new
configurations were found (this occurred typically with distributions in the
temperature range 20 to 30 K). To obtain the highest energy isomers,
additional distributions were generated from Metropolis walks at high
temperatures in the cluster dissociation region (ranging from 60 to 75 K) and
quenched. Again, each configuration was quenched until the relative energy
difference converged to within $10^{-7}$; the final composite file of unique
isomers was then further refined by running another set trajectories until
the relative energy difference converged to within $10^{-12}$.

The efficacy of this method was checked by comparing the results obtained for
a homogeneous thirteen-atom cluster with those reported in the literature.
Hoare and McInnes\cite{Hoare-McInnes} identified 988 unique isomers for
thirteen-atom Lennard-Jones clusters in an extensive study of cluster
morphology based on seeding methods. Tsai and Jordan\cite{TJ_isomers}
identified 1328 isomers using eigenmode methods, ranging in reduced energy
from $-44.3268$ to $-35.0706$. I was able to identify 1167 isomers over the
same range by quenching the J-walker distribution configurations. The
shortfall is most likely due to my having missed some of the higher energy
isomers rather than to having missed lower energy isomers, since the higher
energy isomers are present predominately only in the highest temperature
distributions and thus are more easily missed. The lowest energy isomers form
an increasing fraction of the lower temperature J-walker distributions and so
are encountered frequently enough as the temperature is decreased that it is
likely they were all found; all of the twenty lowest energy isomers listed in
Ref.~\onlinecite{TJ_isomers} where found in the quenched J-walking
distributions. Because it is the lower energy isomers that dominate cluster
behavior in the solid-liquid transition region, and because I am primarily
interested in characterizing cluster solid-liquid transition behavior rather
than obtaining a complete listing of the cluster isomers, the list of isomers
obtained from J-walker distribution quenches was sufficient for my purposes.

More troubling was that some of the quenched configurations were not local
minima, but corresponded to metastable states. Fig.~\ref{Fig:Ar13_Minima}
shows the ten lowest isomers obtained for Ar$_{13}$ from the J-walker
quenches. Two of these were metastable and would eventually relax to a lower
energy configuration when repeatedly quenched again with larger stepsizes.
The $-41.55520$ configuration is in fact the lowest energy first-order saddle
point found by Tsai and Jordan (none of the other first-order saddle points
listed in Ref.~\onlinecite{TJ_isomers} were among the J-walker distribution
quenched configurations, though). To ensure that no metastable configurations
were among the lowest energy binary cluster configurations, the lowest thirty
quenched configurations for each cluster were visually inspected. None were
found.

\section{Results and Discussion}                        \label{Sec:results}
\subsection{Structural properties}
The small size difference between the Ar and Kr atoms suggests that binary
Ar-Kr clusters basically have topological configurations similar to their
homogeneous counterparts, but with many additional permutational isomers
arising from the different arrangements of the Ar and Kr atoms within each
type of topological form. This was indeed the case, as can be seen in
Fig.~\ref{Fig:Minima}, which shows the thirteen lowest energy isomers for
each of the clusters; their energies are listed in Table~\ref{Tbl:Minima}.
These isomers were obtained from quenches of J-walker distribution
configurations and are all readily recognizable as variations of the four
most stable thirteen-atom homogeneous cluster configurations depicted in
Fig.~\ref{Fig:Ar13_Minima}, with the ground state in each case being
icosahedral-like. For those clusters having several permutational isomers,
the segregated isomers have lower energies than the mixed isomers. This is
consistent with the results Lopez and Freeman\cite{Lopez-Freeman} obtained
for Pd$_6$Ni$_7$, where the lowest energy isomer was also completely
segregated (this behavior depended very strongly on the combining rules,
though; increasing the value of $\epsilon_{\mbox{\scriptsize Pd-Ni}}$ by as
little as 2\% above the Berthelot-Lorentz rule led to the mixed isomer having
the lowest energy).

Given the very large energy difference between the ground state icosahedron
and the next three lowest isomers for the homogeneous cluster ($\Delta E_1 =
E_1 - E_0 = 2.85482$ compared to $\Delta E_2 = E_2 - E_1 = 0.02738$ and
$\Delta E_3 = 0.0502$), one might expect that the binary cluster isomers
would likewise form widely spaced groupings, with all the icosahedral-like
permutations in the lowest energy grouping, followed by all the permutational
isomers corresponding to the topological forms having a truncated icosahedral
core with a lone displaced atom, and then all the permutational isomers of
all the other topological forms. The expected ordering of the low-lying
isomers was seen for those clusters consisting mostly of Kr atoms, but not
for the Ar-dominant clusters. For Ar$_{12}$Kr, the other icosahedral-like
configuration having the Kr atom as the central atom was not among the lowest
thirteen isomers (it was actually the 27th lowest isomer found, with an
energy of $-42.05315$). Likewise for Ar$_{11}$Kr$_2$, the Kr-core
icosahedral-like configuration was much higher in energy (ranking 99th with
an energy of $-43.56749$). In each case, the lowest energy isomers all had an
Ar atom as the central atom. Although Ar and Kr atoms are sufficiently
similar in size that the icosahedral-like configurations dominated the lowest
energy isomers, the size difference is large enough that the energy gain
obtained by having a Kr core atom (thereby increasing the number of stronger
Ar-Kr interactions) is more than offset by the energy lost with the increased
separation of the Ar atoms from one another as they are wrapped around the
larger Kr core. As the fraction of Kr atoms increases, this energy difference
decreases, and the Kr-core icosahedral-like isomers became energetically
favorable relative to the non-icosahedral isomers.

The trends in the minimum energy distributions as the fraction of Kr atoms
increases are shown in Fig.~\ref{Fig:Min_Energies}, which depicts the energy
spectrum for each cluster's local minima relative to its global minimum.
Again, the energies were all obtained from J-walker distribution quenches.
The plots have been arranged so that complementary clusters
(Ar$_{13-n}$Kr$_n$ and Ar$_n$Kr$_{13-n}$) appear in the same column. The
densities of the energetically distinct local minima increase as the number
of permutational isomers increases, and for the non-icosahedral isomers
become so great that most of the individual levels cannot be distinguished in
the plots. Those clusters having a greater fraction of Ar atoms have an
energy spectrum similar to that of Ar$_{13}$, which is dominated by its very
large energy gap between the icosahedral ground state and the three closely
spaced truncated icosahedral isomers with their lone displaced atom (and to a
lesser extent by a substantial energy gap between these three isomers and the
subsequent higher energy non-icosahedral isomers). As the fraction of Kr
atoms in the cluster increases from Ar$_{12}$Kr to Ar$_7$Kr$_6$, the number
of Ar-core icosahedral-like permutational isomers increases, forming a small
group of closely spaced low lying energies. Similarly, the number of truncated
icosahedral and non-icosahedral permutations increases rapidly. The energy
spread for the different permutational isomers is large enough that the
higher energy gaps seen in the Ar$_{13}$ spectrum become increasingly filled,
so that by Ar$_{10}$Kr$_3$ the spectrum consists mostly of a small group of
Ar-core icosahedral-like isomers well separated from a large ``band'' of
truncated icosahedral and non-icosahedral isomers; the gap between the two
remains relatively constant as the Kr fraction increases to Ar$_7$Kr$_6$.

The small triangles in the plots each indicate the lowest energy
icosahedral-like isomer found having a Kr central atom. As can be seen in the
plots, the energy difference between these isomers and the ground state
Ar-core isomers decreases steadily as the Kr fraction increases from
Ar$_{12}$Kr, where the isomer energy is much higher than even that of some of
the non-icosahedral isomers, to ArKr$_{12}$, where the energy lies much
closer to the ground state than to the next highest energy isomer. By
Ar$_7$Kr$_6$, the Kr-core icosahedral-like isomers form another small group
that lies intermediate between the low energy Ar-core icosahedral-like group
and the higher energy truncated icosahedral and non-icosahedral band. As a
consequence, the energy spectra for the ArKr$_{12}$ to Ar$_6$Kr$_7$ sequence
of clusters are qualitatively different than those of their Ar$_{12}$Kr to
Ar$_7$Kr$_6$ counterparts. As will be seen later, this qualitative difference
carries over to the heat capacity curves as well.

The decrease in the energy difference between the Kr-core and the Ar-core
icosahedral-like isomers as the Kr fraction increases appears to be nearly
linear across the series. This is a consequence of the mostly linear
dependence of the icosahedral energies as a function of the cluster
composition, as is evident in Fig.~\ref{Fig:E_linear}, which plots the
potential energies of the lowest energy icosahedral-like isomers for the
Ar-core and Kr-core isomers. The overall trend is dominated by the large
difference between the Ar and Kr intermolecular forces --- as the fraction of
Kr atoms increases, so too does the fraction of stronger Ar-Kr and Kr-Kr
interactions. This effect overshadows the smaller, more subtle factors that
influence the cluster structures. These can be made more evident by scaling
out the gross linear dependence on cluster composition. The potential
energies for the Ar-core and Kr-core icosahedral isomers as functions of
cluster composition are shown again in Fig.~\ref{Fig:Min_Energies_By_Comp},
but in energy units scaled by the composition weighted average $\epsilon_{\rm
Avg} = X_{\rm Ar} \epsilon_{\mbox{\scriptsize Ar-Ar}} + X_{\rm Kr}
\epsilon_{\mbox{\scriptsize Kr-Kr}}$, where $X_{\rm Ar} = n_{\rm Ar}/(n_{\rm
Ar} + n_{\rm Kr})$ and $X_{\rm Kr} = 1 - X_{\rm Ar}$. The lower plot shows
the ground state Ar-core icosahedral-like configurations, while the upper
plot shows their corresponding Kr-core isomers. To provide further insight
into the interplay between the atomic size and the intermolecular potential,
I have also included for comparison the energies of the minimized isomers
resulting from having a common atomic size ($\sigma_{\mbox{\scriptsize
Ar-Ar}} = \sigma_{\mbox{\scriptsize Kr-Kr}}$, $\Gamma = 1$), and those
resulting from having a common well depth ($\epsilon_{\mbox{\scriptsize
Ar-Ar}} = \epsilon_{\mbox{\scriptsize Kr-Kr}}$, $\alpha = 1$). For the
Kr-core isomers, the curve is very similar to the
$\epsilon_{\mbox{\scriptsize Ar-Ar}} = \epsilon_{\mbox{\scriptsize Kr-Kr}}$
curve, indicating that it is the different atomic sizes that are the dominant
influence in this case, while for the Ar-core isomers, the curve lies almost
equidistant between the $\epsilon_{\mbox{\scriptsize Ar-Ar}} =
\epsilon_{\mbox{\scriptsize Kr-Kr}}$ and $\sigma_{\mbox{\scriptsize Ar-Ar}} =
\sigma_{\mbox{\scriptsize Kr-Kr}}$ curves, implying that neither factor
predominates (it is slightly closer to the $\sigma_{\mbox{\scriptsize Ar-Ar}}
= \sigma_{\mbox{\scriptsize Kr-Kr}}$ curve).

The two sets of curves are complementary in the sense that the largest
deviations from the homogeneous cluster values occur in each case for the
isomers having a core atom of one component surrounded by twelve atoms of the
other component, while the smallest deviations occur for the isomers having a
single atom of one component occupying an exterior site on the twelve-atom
sub-cluster of the other component. Also, the directions of the deviations
are reversed. For the ground state Ar-core isomers, setting
$\sigma_{\mbox{\scriptsize Ar-Ar}} = \sigma_{\mbox{\scriptsize Kr-Kr}}$
raises the scaled energy, while setting $\epsilon_{\mbox{\scriptsize Ar-Ar}}
= \epsilon_{\mbox{\scriptsize Kr-Kr}}$ lowers the scaled energy; the opposite
is true for the Kr-core isomers.

Setting the Ar and Kr sizes equal and minimizing the resulting configurations
isolates the effects of the intermolecular potential on the minimum energies
of the two sets of isomers. For the Kr-core isomers, setting
$\sigma_{\mbox{\scriptsize Ar-Ar}} = \sigma_{\mbox{\scriptsize Kr-Kr}}$
effectively replaces the larger core atom with a smaller one, allowing the
distended outer atoms to pack more tightly and thus lower the overall energy.
The effect is most pronounced for Ar$_{12}$Kr since all twelve Ar atoms can
move in a substantial amount; for ArKr$_{12}$, setting
$\sigma_{\mbox{\scriptsize Ar-Ar}} = \sigma_{\mbox{\scriptsize Kr-Kr}}$ is
equivalent to replacing the smaller exterior Ar atom with a larger one, which
has only a minor effect. The scaled energies are all less than that of
homogeneous Ar$_{13}$ because of the contributions from the stronger Ar-Kr
and even stronger Kr-Kr interactions (much of the contribution from the Kr-Kr
interactions has been incorporated into the energy scaling, but since the
scaling is based only on the mole fraction and not on the actual distribution
of the different interactions, small variations remain). For the Ar-core
isomers, the opposite occurs; setting $\sigma_{\mbox{\scriptsize Ar-Ar}} =
\sigma_{\mbox{\scriptsize Kr-Kr}}$ effectively replaces the smaller core atom
with a larger one, which moves the outer atoms out more and raises the
overall energy. Analogously, the effect is most pronounced for ArKr$_{12}$
since all twelve Kr atoms are distended a substantial amount, while for
Ar$_{12}$Kr, only the larger exterior atom is replaced by a smaller one. The
scaled energies for the Ar-core isomers are all greater than that of
Ar$_{13}$ because the central Ar atom decreases the number of Kr-Kr
interactions relative to the weaker Ar-Kr and Ar-Ar interactions. These
trends imply that the Kr-core icosahedral-like isomers would be the
ground-state isomers if the two component atoms were actually the same size,
and so there exists a critical size ratio for a given pair of binary cluster
components where the isomers having the smaller atom (with the shallower well
depth) at the core attain lower energies than those having the larger atom
(with the deeper well depth) at the core.

Setting the Ar and Kr well depths equal and minimizing the configurations
isolates the effects of atomic size on the minimum energies of the two sets
of isomers. For the Ar-core isomers, the scaled energy is lowered in each
case, with the largest decrease occurring for ArKr$_{12}$ and the smallest
for Ar$_{12}$Kr. The effects can be more easily understood from the
perspective of modifying the corresponding homogeneous clusters rather than
from the modifications to the binary clusters themselves. In each of these
two cases, setting $\epsilon_{\mbox{\scriptsize Ar-Ar}} =
\epsilon_{\mbox{\scriptsize Kr-Kr}}$ is equivalent to making one of the atoms
of a homogeneous Ar$_{13}$ or Kr$_{13}$ cluster a different size. For
ArKr$_{12}$, this atom would be the core atom, which effectively shrinks,
again allowing all twelve of the exterior atoms to pack in more tightly and
lower the energy, while for Ar$_{12}$Kr, this atom would be one of the
exterior atoms, which effectively expands. This has a much smaller effect,
though, since the atom is on the exterior. In a similar fashion for the
Kr-core isomers, setting $\epsilon_{\mbox{\scriptsize Ar-Ar}} =
\epsilon_{\mbox{\scriptsize Kr-Kr}}$ is again equivalent to changing the
sizes of some of the atoms of a homogeneous thirteen-atom cluster. For
Ar$_{12}$Kr, the atom now effected would be the core atom, which becomes
bigger, thus pushing all twelve of the outer Ar atoms outward and raising the
overall energy a large amount, while for ArKr$_{12}$, the atom would be only
one of the exterior atoms, which again would increase the energy only
slightly. The energy increases relative to the homogeneous clusters are much
larger for the Kr-core isomers than are the corresponding energy decreases
for the Ar-core isomers; because of the repulsive steric constraints, the
exterior Kr atoms in the Ar-core isomers are more restricted in their ability
to collapse around the smaller Ar atom than are the exterior Ar atoms in the
Kr-core isomers in their ability to distend outward from the larger Kr atom.

\subsection{Thermodynamic properties}
The qualitative differences between the structural properties of the
predominately Ar clusters and those of the predominately Kr clusters are
found in their thermodynamic properties as well. Fig.~\ref{Fig:Cv} shows the
heat capacity for each binary cluster as a function of temperature. The
J-walking curves in each case are the averages obtained from combining the
results of the five individual J-walking runs, as described in
Section~\ref{Sec:theory}. Representative standard deviations have not been
included to avoid cluttering the figures, but they were consistent with the
noise levels in the curves. While the Ar$_{12}$Kr heat capacity curve is very
similar to the Ar$_{13}$ curve obtained previously\cite{J-walker,Magic_Cv}
with a dominating peak in the solid-liquid transition region at 34 K, the
ArKr$_{12}$ curve is very much different, having a well separated second peak
at 18 K. The heat capacity curve for each cluster in the series is
characterized by a very large peak in the solid-liquid transition region at
about 35 to 45 K, implying that each cluster still retains its magic number
status, but significant changes are evident across the series from
Ar$_{12}$Kr to ArKr$_{12}$. Analogous to the Ar$_{13}$ case, the large size
of the solid-liquid transition peak in each binary cluster is due to the
correspondingly large energy difference between the ground-state
icosahedral-like isomers and the higher energy non-icosahedral isomers, as
well as to the large barriers to interconversion between the isomers. The
variations in the peaks as the number of Kr atoms increases, and the ultimate
appearance of the second peak in ArKr$_{12}$, are consequences mostly of
energy differences occurring between the Ar-core and Kr-core isomers. The
solid-liquid transition heat capacity peaks for those clusters having a
substantial Ar fraction (Ar$_{12}$Kr to Ar$_5$Kr$_8$) are mostly alike,
showing only some minor differences in the peak heights, as well as a slight
shifting of the peak position to higher temperatures as the number of Kr
atoms increases. For the Kr-dominant clusters (Ar$_4$Kr$_9$ to ArKr$_{12}$),
though, a smaller peak can be seen developing on the low temperature side of
the solid-liquid transition peak, shifting downward in temperature with
increasing Kr numbers, until becoming well separated at ArKr$_{12}$. As was
seen in Fig.~\ref{Fig:Min_Energies}, the energy difference between the
Ar-core and Kr-core icosahedral-like isomers decreases steadily and rapidly
across the series. The Kr-core icosahedral-like isomers first separate from
the large non-icosahedral band at Ar$_7$Kr$_6$, become approximately
equidistant between the ground-state Ar-core icosahedral-like group and the
non-icosahedral band by Ar$_4$Kr$_9$, and finally attain their nearest
approach to the ground state at ArKr$_{12}$. Thus for the predominately Ar
clusters, the Kr-core icosahedral-like isomers are too high in energy to be
accessible at temperatures below the solid-liquid transition region, but for
the clusters Ar$_3$Kr$_{10}$ to ArKr$_{12}$, the Kr-core icosahedral-like
isomers are sufficiently low lying in energy that they contribute to the heat
capacity at temperatures well below the solid-liquid transition region.

The standard Metropolis heat capacity curves for each cluster are also
included in Fig.~\ref{Fig:Cv}. These are in good agreement with the J-walking
curves only for the predominantly Ar clusters. For Ar$_{12}$Kr to
Ar$_9$Kr$_4$, the Metropolis solid-liquid transition peaks are about 5 to
10\% too low, but otherwise compare well with their J-walking counterparts.
This is comparable to the size of the discrepancies found previously for
Ar$_{13}$.\cite{J-walker,Magic_Cv} For Ar$_8$Kr$_5$ to Ar$_3$Kr$_{10}$,
though, there are substantial discrepancies between the Metropolis and
J-walking curves evident on the low temperature side of the solid-liquid
transition peak, and for Ar$_2$Kr$_{11}$ and ArKr$_{12}$, the low temperature
shoulder and peak, respectively, are completely absent in the Metropolis
curves. The deficiency of the standard Metropolis algorithm in application to
binary clusters in this temperature regime can also be seen by comparing the
results to those obtained using the atom-exchange method, which have been
included for the clusters Ar$_6$Kr$_7$ to ArKr$_{12}$. In each case, the
atom-exchange results are in good agreement with the J-walker results, with
the exception of the transition peak heights again being slightly too low.
Even the smaller low temperature peak in ArKr$_{12}$ was well reproduced with
the atom-exchange method. Because the atom-exchange method is just the
standard Metropolis algorithm augmented with an occasional swapping of atoms
from the different components, it is clear that the Kr-core icosahedral
isomers are not being accessed dynamically during the course of the
Metropolis walks at these temperatures, but only via the exchange mechanism
(or from the stored distributions in the case of the J-walking runs).

The slight energy differences between the various Ar-core icosahedral-like
permutational isomers result in the very small mixing-anomaly peaks that
occur at very low temperatures (about 1 to 5 K) in Fig.~\ref{Fig:Cv}. In
general, the size of each peak is a function of the number of Ar-core
icosahedral-like isomers present --- Ar$_{12}$Kr, Ar$_2$Kr$_{11}$ and
ArKr$_{12}$ each have only one Ar-core icosahedral-like isomer and thus do
not exhibit this low temperature mixing-anomaly peak (although
Ar$_2$Kr$_{11}$ and ArKr$_{12}$ show mixing between the ground state Ar-core
isomer and the Kr-core isomers), Ar$_{11}$Kr$_2$ and Ar$_3$Kr$_{10}$ each
have just three closely spaced Ar-core icosahedral-like permutational isomers
that can contribute to the heat capacity and so have barely discernible
mixing-anomaly peaks, while all the remaining clusters from Ar$_{10}$Kr$_3$
to Ar$_4$Kr$_9$ have enough low lying Ar-core icosahedral-like permutational
isomers that the mixing-anomaly peak in each case is clearly noticeable. The
peak heights are all very small, less than five reduced units from the
baseline. Considering that the standard Metropolis method was unable to
handle the transitions to the low lying Kr-core icosahedral-like isomers in
Ar$_2$Kr$_{11}$ and ArKr$_{12}$, it is not surprising then that the method
was also unable to reproduce the mixing-anomaly peaks. The atom-exchange
results show some smaller peaks, but they are not in good agreement with the
J-walking results. This can be seen better in Fig.~\ref{Fig:Low_T_Cv}, which
shows the low-temperature parts of the heat capacity curves obtained from
each of the three methods in greater detail for all the clusters having more
than one Ar-core icosahedral-like isomer. The atom-exchange curves are all in
good agreement with the J-walking curves on the high temperature tails of the
peaks, but each drops off rapidly, joining the Metropolis curves well before
the peak temperature. This is a consequence of the atom-exchange method
failing as the temperature becomes too low. At such low temperatures, most of
the Metropolis configurations have energies only slightly above their nearest
local minimum so that the trial configurations most likely obtained by
exchanging an exterior Ar atom with an exterior Kr atom will be a strained
configuration having too high an energy to be accepted. The dependence of the
peak size on the number of Ar-core isomers can also be seen more clearly:
Ar$_7$Kr$_6$ has the most isomers with eighteen, and has the largest peak;
Ar$_8$Kr$_5$ and Ar$_6$Kr$_7$ with twelve isomers each, and Ar$_9$Kr$_4$ and
Ar$_5$Kr$_8$ with ten each, have similarly sized peaks; Ar$_{10}$Kr$_3$ and
Ar$_4$Kr$_9$ with five isomers each have much smaller peaks, and
Ar$_{11}$Kr$_2$ and Ar$_3$Kr$_{10}$ with only three isomers each have the
smallest peaks (which are actually only shoulders). The peak widths
also depend on the energy spread between the different isomers. For example
Ar$_{10}$Kr$_3$ has a smaller energy spread than does Ar$_4$Kr$_9$ ($\Delta E
= E_4 - E_0 = 0.12115$ compared to 0.16146), and thus it has a narrower peak.

Lopez and Freeman\cite{Lopez-Freeman} estimated the entropy change associated
with the low temperature isomerization transition for Pd$_6$Ni$_7$ by
numerically integrating their heat capacity data according to the expression
\begin{equation}
    \Delta S_t = \int_0^{T_1} \frac{\Delta C_V\,dT}{T},   \label{Eq:Delta_S}
\end{equation}
where $T_1$ was the temperature just above the mixing anomaly, and $\Delta
C_V$ was the difference between the J-walking and Metropolis heat capacities
(the assumption being that the J-walking results reflected complete access to
all the permutational isomers over the temperature range, while the system
was stuck in its lowest-energy configuration during the Metropolis runs so
that none of the permutational isomers were accessed). They obtained a value
of $\Delta S_t /k_B = 4.8$. I have followed Lopez and Freeman and have
calculated the entropy changes from the heat capacity data shown in
Fig.~\ref{Fig:Low_T_Cv} by numerically integrating Eq.~\ref{Eq:Delta_S}. The
results are plotted in Fig.~\ref{Fig:Entropy}. The entropy changes mostly
support the qualitative observations given earlier, that the peak sizes are
related to the number of closely spaced Ar-core icosahedral-like
permutational isomers, except for the curious inversion of the values for
Ar$_8$Kr$_5$ and Ar$_6$Kr$_7$ (both with twelve isomers), which are slightly
lower than those for Ar$_9$Kr$_4$ and Ar$_5$Kr$_8$ (both with ten isomers).
The curve is nearly symmetrical about Ar$_7$Kr$_6$, consistent with the
symmetry in the numbers of Ar-core isomers. The maximum value (4.4) occurring
for Ar$_7$Kr$_6$ is similar to Lopez and Freeman's value for the analogous
metal cluster Pd$_6$Ni$_7$.

The shifting of the solid-liquid transition peaks to higher temperatures as
the Kr fraction increases, as well as the variation in the peak heights, is
presented in more detail in Fig.~\ref{Fig:Cv_Peaks}; the peak heights and
temperatures are also listed in Table~\ref{Tbl:Cv_peaks}. These values were
obtained in each case from the averaged J-walker curves shown in
Fig.~\ref{Fig:Cv} The curves were smoothed and then interpolated in the peak
vicinities to obtain finer mesh sizes. The peak parameters were then found
simply by searching the interpolated data. Some minor variations in the peak
parameter values thus obtained occurred when different smoothing parameters
were used,\cite{Sav_Gol} but these were much smaller than the uncertainties
obtained from averaging the individual J-walker runs, and so a single common
set of smoothing parameters was used for each case, with the uncertainties
in the peak heights then estimated from the average standard deviations
of the points in the vicinity of each peak. The uncertainties in the peak
height were all less than 1\%, and the uncertainties in the peak temperature
less than 0.5\%. The values for Ar$_{13}$ ($118.5 \pm 0.6$ at $34.05 \pm
0.06$~K) and scaled Kr$_{13}$ ($118.0 \pm 0.6$ at $34.05 \pm 0.07$~K) are in
good agreement with the values I obtained previously ($117.7 \pm 0.2$ at
$34.33 \pm 0.08$~K),\cite{Magic_Cv} and with the values obtained by Tsai and
Jordan (117.1 at 34.15~K).\cite{Tsai-Jordan,Kr-note} Table~\ref{Tbl:Cv_peaks}
also lists the mixing-anomaly peak parameters for those clusters having
distinct peaks.

As is evident in Fig.~\ref{Fig:Cv_Peaks}, there is a nearly linear increase
in the peak temperature as a function of cluster composition from Ar$_{13}$
(34.05~K) to Ar$_7$Kr$_6$ (37.05~K), followed by another nearly linear, but
more rapid increase in the peak temperature for the clusters Ar$_6$Kr$_7$
(37.63~K) to Kr$_{13}$ (46.77~K). The peak heights show a more interesting
variation. They increase initially as the fraction of Kr atoms increases from
Ar$_{13}$, reaching a maximum at Ar$_9$Kr$_4$. They then continue decreasing
until reaching a minimum at Ar$_3$Kr$_{10}$, before rising again until they
reach their original height again at Kr$_{13}$. This complementary behavior
is consistent with the differences in the cluster minimum energies between
the Ar-dominant and Kr-dominant clusters. It should be kept in mind, though,
that these variations are relatively small (the difference between the
maximum and minimum peak heights is only about 12\% of the Ar$_{13}$ peak
height) and that the heat capacity peak heights are in fact remarkably
similar considering the degree of change seen in the cluster structural
properties as functions of composition.

The potential energy curve for each cluster as a function of temperature is
shown in Fig.~\ref{Fig:Caloric}. Except for the ArKr$_{12}$ curve, the curves
all appear very similar in shape and show roughly uniform spacing. This
again is a consequence of the mostly linear dependence of the isomer energies
on the cluster composition. The low temperature part of the curves is
dominated by the energetics of the Ar-core icosahedral-like isomers and thus
exhibit much the same linear dependence with cluster composition as was seen
in Fig.~\ref{Fig:E_linear}, while the high temperature part of the curves is
dominated by various dissimilar liquid-like structures that are well mixed.
In this regime, the clusters are sufficiently distended that the
intermolecular interactions predominate over the size constraints, resulting
in curves having almost uniform displacements.

\subsection{Quench results}
The temperature dependence of the binary cluster isomerizations can also be
inferred from the quench studies. Figs.~\ref{Fig:Ar_Quench} and
\ref{Fig:Kr_Quench} show the results obtained using J-walking for the
Ar-predominate and Kr-predominate clusters, respectively; quench results for
Ar$_{13}$ have also been included in Fig.~\ref{Fig:Ar_Quench} for comparison.
The large energy gap between the Ar$_{13}$ ground state icosahedral isomer and
the next lowest energy isomers is reflected in the quench curves --- all the
configurations quenched to the icosahedral isomer, up to a temperature of
25~K, and half the configurations quenched to the icosahedral isomer at the
heat capacity peak temperature of 34~K. Even at a temperature of 45~K, which
is well into the liquid region, about 10\% of the quenches were to the lowest
energy isomer. The next three lowest energy isomers (the truncated
icosahedral isomers having a lone displaced atom located on one of the
icosahedral faces) are closely spaced in energy and well separated from the
next highest energy isomer. Their quench curves are likewise very similar,
forming a distinct group of curves slowly rising from zero at about 28~K.
These three isomers and the lowest energy isomer dominate the cluster
dynamics in the transition region; at the heat capacity peak temperature,
they comprise 80\% of the quenched configurations. Given the similarity
between the Ar$_{13}$ and Ar$_{12}$Kr isomer energy spectra shown in
Fig.~\ref{Fig:Min_Energies}, it is not surprising that their quench curves
are also similar. The group of three closely spaced Ar$_{13}$ truncated
icosahedral isomers becomes a slightly broader group of twenty similar
Ar$_{12}$Kr permutational isomers, but the quench profile for this group of
isomers is similar to the profile for the three Ar$_{13}$ isomers.

Ar$_{11}$Kr$_2$ has three Ar-core icosahedral-like permutational isomers
closely spaced in energy that can be accessed at very low temperatures, and
this is evident in the quench curves. The curves are roughly constant over
the domain 15~K to 30~K, with the two lowest energy isomers being
predominate, comprising about 90\% of the quenched configurations. The
apparent deficiency of quenches for the third isomer can be explained simply
in terms of the number of equivalent configurations. As shown in
Fig.~\ref{Fig:Minima}, this isomer has the two Kr atoms occupying the
axial positions of the icosahedron. The number of ways the two Kr and eleven
Ar atoms can be combined to form this icosahedral isomer is much smaller than
the number of ways the other two isomers can be formed. The behavior at the
heat capacity peak temperature is similar to that of Ar$_{13}$ and
Ar$_{12}$Kr, with the number of Ar-core icosahedral quenches decreasing
quickly and the number of non-icosahedral quenches rising rapidly. The
Ar$_{10}$Kr$_3$ curves are similar to the Ar$_{11}$Kr$_2$ curves, except that
Ar$_{10}$Kr$_3$ has five closely spaced Ar-core icosahedral-like
permutational isomers. As the temperature increases from zero, each isomer is
accessed in order according to its energy, but over the range 10~K to 30~K
where the curves show plateaus, the levels are determined by the
combinatorics. Thus the ground state isomer having its three Kr atoms
occupying adjacent icosahedral sites has the second lowest number of quenches
over this region.

The plots for the clusters Ar$_9$Kr$_4$ to Ar$_5$Kr$_8$ were mostly similar
to the Ar$_{10}$Kr$_3$ plot, except that they had many more low lying
Ar-core icosahedral-like permutational isomers, and so these plots have not
been included. For the Kr-predominate clusters, though, the relatively low
lying Kr-core icosahedral-like permutational isomers can also be seen to
effect the quench curves. Like Ar$_{10}$Kr$_3$, Ar$_4$Kr$_9$ has five closely
spaced Ar-core icosahedral-like permutational isomers. It also has ten
closely spaced Kr-core icosahedral-like permutational isomers having energies
roughly equidistant between the ground state isomer and the non-icosahedral
isomers. As can be seen in Fig.~\ref{Fig:Minima}, the arrangements of the
three non-core Ar atoms in each of the five Ar$_4$Kr$_9$ Ar-core
icosahedral-like isomers is the same as the arrangements of the three Kr
atoms in each of the corresponding Ar$_{10}$Kr$_3$ isomers, and so the
low-temperature parts of their quench curves are similar, with the isomers
showing the same ranking in the number of quenched configurations in the
plateau region (and nearly the same levels). The quenched Ar$_4$Kr$_9$
Kr-core icosahedral-like permutational isomers begin to appear at about 20~K,
which is consistent with the broadening of the solid-liquid transition region
heat capacity peak seen in Fig.~\ref{Fig:Cv}. The non-icosahedral isomers
first appear at about 25~K, similar to the other clusters.

Analogously, both Ar$_3$Kr$_{10}$ and Ar$_{11}$Kr$_2$ have three low lying
Ar-core icosahedral-like permutational isomers, with the two exterior Ar
atoms in Ar$_3$Kr$_{10}$ having the same three arrangements as the two Kr
atoms in Ar$_{11}$Kr$_2$. Again the quench curves for the three
Ar$_3$Kr$_{10}$ isomers are very similar to the corresponding three
Ar$_{11}$Kr$_2$ isomers at temperatures below 20~K, but the Ar$_3$Kr$_{10}$
curves begin to drop off at this point as the five Kr-core icosahedral-like
permutational isomers begin to be accessed. The Kr-core isomers are close
enough in energy to the Ar-core icosahedral-like isomers that their quench
curves can be seen to be well separated from the non-icosahedral isomers,
unlike the case for the other binary clusters having a greater Ar fraction.
This trend continues with increasing Kr fraction. Ar$_2$Kr$_{11}$ has only
one Ar-core icosahedral-like isomer, and so the three Kr-core
icosahedral-like isomers are the next isomers accessed, beginning at about
15~K. Finally, with ArKr$_{12}$, the Kr-core icosahedral-like isomer is so
much closer in energy to the Ar-core icosahedral-like ground state than it
is to the non-icosahedral isomers that 90\% of the quenches at 35~K, where the
non-icosahedral isomers begin to be accessed, are to the Kr-core isomer. Note
that while both isomers each have half of the quenched configurations at 20~K,
a temperature near the heat capacity peak temperature, this does not imply
that there is a dynamic equilibrium between the two. The barrier height for
moving an Ar atom from the core to an exterior site while a Kr atom moves
into the core is too high to overcome at these temperatures, as is implied by
the fact that the peak is not reproduced at all using the standard Metropolis
method.

\section{Conclusion}                                \label{Sec:conclusion}
Binary clusters whose component atoms are not too dissimilar retain many of
the features of their simpler homogeneous counterparts, but have additional
parameters such as the cluster composition and mixing rules that can be
``tuned'' to modify their properties and provide more insight into their
physical behavior. Much of the physical behavior of thirteen-atom Ar-Kr
clusters was very similar to that of Ar$_{13}$, especially for those clusters
having a low Kr fraction, but several changes that occurred as the Kr fraction
increased resulted in the predominantly Kr clusters being quite different.
These changes were not so drastic, though, that they radically altered the
physical properties --- even ArKr$_{12}$, which was the most dissimilar of
the set, still retained the icosahedral-like ground-state and the
magic-number behavior characteristic of homogeneous thirteen-atom clusters.
Replacing Ar atoms in Ar$_{13}$ by Kr atoms may have modified the potential
energy hypersurface considerably, and in the case of the predominately Kr
clusters led to several isomers having energies within the large gap between
the icosahedral global minimum and the non-icosahedral local minima, but the
hypersurface still retained the characteristic deep wells and large barriers
associated with the icosahedral-like isomers that dominate the solid-liquid
transition behavior.

The changes in the Ar-Kr clusters' heat capacity behavior as the Kr fraction
increased were primarily due to the Ar-core and Kr-core icosahedral-like
permutational isomers. Although the Ar and Kr sizes are similar enough that
all the binary clusters in the series had icosahedral-like isomers as their
lowest energy configurations, the size difference is large enough that the
central atom in each case was an Ar atom, instead of a Kr atom, which would
have been the case had their sizes been equal. For the predominately Ar
clusters, the Kr-core icosahedral-like isomers were too high in energy to
have had much effect on the clusters' low-temperature behavior, but as the Kr
fraction increased across the series, the Kr-core icosahedral-like isomers
rapidly dropped in energy, so that for the predominately Kr clusters, they
began to dominate the clusters' low-temperature behavior, eventually forming
a distinct second, lower temperature, heat capacity peak. The Ar-core
icosahedral-like permutational isomers manifested themselves as the very
small mixing-anomaly peaks that occurred at very low temperatures. These peaks
had sizes that depended primarily on the number of Ar-core permutational
isomers, since the energy differences between these isomers were quite small.
The mixing-anomaly peak was largest for Ar$_7$Kr$_6$, which has the most
Ar-core permutational isomers.

Since the mixing-anomaly heat capacity peaks are due to the presence of
several closely spaced isomers having energies only slightly above that of the
lowest-energy isomer, one may wonder why similar, very small, low-temperature
peaks have not been previously observed in homogeneous clusters likewise
having low lying isomers whose energies lie slightly above their ground-state
isomer. For example, Ar$_{17}$ has a low lying isomer with an energy of
$-61.307$ that lies only slightly above the ground-state isomer energy of
$-61.318$.\cite{FD} In my previous study of Lennard-Jones cluster heat
capacities,\cite{Magic_Cv} neither the $N = 17$ curve nor any of the other
heat capacity curves I calculated for cluster sizes ranging from four to
twenty-four atoms had such peaks, although curves for $N = 18$ and 21 showed
very small, low temperature, ``bumps'' that might be suitable candidates.
Since the major emphasis in those calculations was the characterization of
magic-number behavior in the solid-liquid transition region, I did not extend
the J-walking distributions down to the very low temperatures that I did in
this study,\cite{low-T-JW} nor did I perform quench calculations to determine
the energetics of the low-lying isomers. Thus it is possible some of those
clusters actually have very small, low temperature, peaks that were simply
missed, although it is not likely for most since their small sizes imply they
do not have many isomers near the ground-state. As the cluster size
increases, though, the number of isomers increases rapidly, and so some
larger clusters could be expected to also possess small low-temperature peaks
in their heat capacity curves. Preliminary heat capacity calculations I have
done on medium sized Lennard-Jones clusters have found such peaks for $N =
31$, 32 and 33, but not for $N = 25$ to 30 (I have not yet done calculations
for $N > 33$). $N = 31$ corresponds to a critical size in Lennard-Jones
cluster structures, in the sense that it marks the crossover point for their
preferred sublattice topologies. In his study of the structure and binding of
Lennard-Jones clusters for $13 \leq N \leq 147$, Northby\cite{Northby}
distinguished between two types of sublattice structures: the ``IC''
sublattice, which consists of all the sites that will comprise the outer
shell of the next complete Mackay icosahedron, and the ``FC'' sublattice,
which consists of those tetrahedrally bonded face sites that lie at stacking
fault locations relative to the first lattice, together with the vertex
sites. For $N < 31$, each of the cluster's lowest energy isomers had an FC
topology, while for $31 \leq N \leq 55$, each had an IC topology. The small
heat capacity peaks for $N = 31$, 32, and 33 appear therefore to reflect a
``phase'' transition between the two different types of structures.

The calculations presented here have demonstrated again that the standard
Metropolis algorithm is inadequate for dealing with the low-temperature
behavior of heterogeneous clusters. Augmenting the Metropolis method with the
atom-exchange method improved its accuracy greatly, and so should be done
routinely on multicomponent simulations where J-walking is not done, to
provide reasonable reliability in this regime. But even the atom-exchange
method does not reduce the problems due to quasiergodicity to the extent that
the J-walking method does. The J-walking calculations for binary clusters
were more complicated than those done on similar homogeneous clusters because
of the greater possibility for systematic errors corrupting the J-walker
distributions, which necessitated the use of multiple trials from
independently generated distributions to ensure that such errors were not
present. However, this computational overhead was mitigated by the shorter
walk lengths required for J-walking to achieve the desired level of accuracy
compared to the Metropolis method; for example, for Ar$_8$Kr$_5$, five
J-walker runs of $10^6$ passes each provided a level of accuracy in the
solid-liquid transition region that required runs totalling $10^8$ passes for
the Metropolis method to match.

Calculations on Ne-Ar clusters similar to the ones reported here for Ar-Kr
clusters are nearing completion and will be submitted for publication
shortly. The interaction parameters for Ne and Ar are much more dissimilar
than are the Ar and Kr interaction parameters ($\epsilon_{\mbox{\scriptsize
Ne-Ne}} / \epsilon_{\mbox{\scriptsize Ar-Ar}} = 0.2982$ compared to
$\epsilon_{\mbox{\scriptsize Ar-Ar}} / \epsilon_{\mbox{\scriptsize Kr-Kr}} =
0.7280$). More importantly though, the Ne and Ar atomic sizes are considerably
more dissimilar than are the Ar and Kr sizes ($\sigma_{\mbox{\scriptsize
Ne-Ne}} / \sigma_{\mbox{\scriptsize Ar-Ar}} = 0.8073$ compared to
$\sigma_{\mbox{\scriptsize Ar-Ar}} / \sigma_{\mbox{\scriptsize Kr-Kr}} =
0.8897$). This size differential is sufficiently large that not all the Ne-Ar
clusters retain the icosahedral-like symmetry for their ground-state
configurations, and a much greater diversity of structures is obtained. Not
surprisingly, this leads to very much different behavior in the Ne-Ar
clusters' heat capacities.

\acknowledgments
The support of the Natural Sciences and Engineering Research Council of
Canada (NSERC) for this research is gratefully acknowledged. I thank the
University of Lethbridge Computing Services for generously providing me the
use of their workstations for some of the calculations reported, and the
University of Waterloo for the use of their facilities. I also thank David
L. Freeman for helpful discussions.


\begin{figure}
\caption{Heat capacity curves for Ar$_8$Kr$_5$ clusters. The large peak
corresponds to the solid-liquid transition region, while the small
low-temperature peak is due to the icosahedral-like permutational isomers.
The five solid curves were obtained from separate J-walking runs, each using
a different set of J-walker distributions, initially generated at 52~K. The
data for each temperature were obtained from $10^6$ total passes. The
differences between the curves are comparable to the noise levels inherent in
each, indicating that the systematic errors associated with each J-walker
distribution were sufficiently small. For comparison, the dotted curves show
the results obtained from similar J-walker runs using sets of J-walker
distributions initially generated at 70~K, a temperature high enough for the
cluster to be completely dissociated. Systematic errors in the high
temperature J-walker distributions can be seen to have not only slightly
effected the solid-liquid transition peak, but to have also propagated
through the subsequent distributions to substantially effect the low
temperature peak. Also included is the dashed curve obtained from standard
Metropolis runs having $10^7$ total passes of data accumulation at each
temperature. The curve was begun at the lowest temperature using the lowest
energy isomer as the initial configuration, with the final configuration at
each temperature then used as the initial configuration for the next
temperature. Substantial discrepancies due to quasiergodicity are evident in
the solid-liquid transition region, and the low-temperature mixing-anomaly
peak is completely absent. The open circles, representing Metropolis results
obtained using $10^8$ total passes, are in agreement with the J-walker
results. Augmenting standard Metropolis with the atom-exchange algorithm does
improve its low temperature performance, as can be seen by the long-dashed
curve, which shows a small mixing-anomaly peak. However, substantial
discrepancies still remain.
\label{Fig:five_trials}}
\end{figure}

\begin{figure}
\caption{Lowest energy configurations for Ar$_{13}$ clusters obtained from
quenches of clusters stored in the J-walker distribution files. The energies
are in reduced units, $E^* = E/\epsilon_{\mbox{\scriptsize Ar-Ar}}$. Except
for the two metastable structures (indicated by the parentheses) these
configurations are identical to the lowest-lying isomers given in
Refs.~\protect\onlinecite{Hoare-McInnes} and \protect\onlinecite{TJ_isomers}.
The four lowest energy stable isomers are the topological forms for all the
lowest energy Ar-Kr isomers shown in Fig.~\protect\ref{Fig:Minima}.
\label{Fig:Ar13_Minima}}
\end{figure}

\begin{figure}
\caption{The thirteen lowest energy isomers found for Ar$_{13-n}$Kr$_n$
clusters ($1 \leq n \leq 12$), in order of increasing energy; their energies
are listed in Table~\protect\ref{Tbl:Minima}. The isomers were obtained from
quenches of configurations stored in J-walker distribution files. In each
case, the ground state isomer is an icosahedral-like configuration having an
Ar atom as the central atom. For those clusters having several permutational
isomers, the segregated isomers have lower energies than the mixed isomers.
\label{Fig:Minima}}
\end{figure}

\begin{figure}
\caption{Local potential energy minima (in reduced units) for
Ar$_{13-n}$Kr$_n$ clusters ($0 \leq n \leq 13$). In each case, the zero
energy level corresponds to the global minimum energy. The triangles mark the
lowest-energy icosahedral-like permutational isomers having a Kr atom as the
central atom.
\label{Fig:Min_Energies}}
\end{figure}

\begin{figure}
\caption{Binary cluster potential energies for the lowest energy
icosahedral-like isomers, as functions of composition. The filled circles
represent the lowest energy isomers having a central Ar atom, while the open
circles represent the lowest energy isomers having a central Kr atom
(indicated by the triangles in Fig.~\protect\ref{Fig:Min_Energies}).
\label{Fig:E_linear}}
\end{figure}

\begin{figure}
\caption{Binary cluster minimum energies as a function of composition.
Energies are scaled by the composition weighted average $\epsilon_{\rm Avg} =
X_{\rm Ar} \epsilon_{\mbox{\scriptsize Ar-Ar}} + X_{\rm Kr}
\epsilon_{\mbox{\scriptsize Kr-Kr}}$, where $X_{\rm Ar} = n_{\rm Ar}/(n_{\rm
Ar} + n_{\rm Kr})$ and $X_{\rm Kr} = 1 - X_{\rm Ar}$. Ground state isomers
are indicated in the lower plot by the filled circles, while the
corresponding lowest-energy Kr-core isomers are indicated in the upper plot
by the open circles. Also shown for comparison are the results for the
artificial cases where the two components have same atomic sizes or the same
well depths; the triangles indicate the minimum energies obtained when
$\sigma_{\mbox{\scriptsize Ar-Ar}}$ was set equal to
$\sigma_{\mbox{\scriptsize Kr-Kr}}$ ($\Gamma = 1$) and each configuration
reminimized, while the squares indicate the minimum energies that resulted
when $\epsilon_{\mbox{\scriptsize Ar-Ar}}$ was set equal to
$\epsilon_{\mbox{\scriptsize Kr-Kr}}$ ($\alpha = 1$) and the isomers
reminimized.
\label{Fig:Min_Energies_By_Comp}}
\end{figure}

\begin{figure}
\caption{Heat capacities for the binary clusters Ar$_{13-n}$Kr$_n$ ($1 \leq n
\leq 12$). In each case, the thick curve represents the J-walking results
and the thin curve the standard Metropolis results; the open circles
represent the results obtained using Metropolis Monte Carlo augmented with
the atom-exchange algorithm. Increasingly larger discrepancies between the
Metropolis and J-walking results over much of the temperature ranges are
evident as the mole fraction of Kr increases. The agreement between the
J-walking results and those obtained using the atom-exchange method is mostly
very good except at very low temperatures. Expanded views of the low
temperature region can be seen in Fig.~\protect\ref{Fig:Low_T_Cv}.
\label{Fig:Cv}}
\end{figure}

\begin{figure}
\caption{Low temperature heat capacity curves for those binary Ar-Kr clusters
exhibiting low temperature mixing-anomaly peaks. In each case, the thick
curve represents the J-walking results and the thin curve the standard
Metropolis results; the dotted curves represent the results obtained using
Metropolis Monte Carlo augmented with the atom-exchange algorithm. The peaks
are due to the low lying Ar-core icosahedral-like permutational isomers, and
tend to be largest for those clusters having the largest number of such
isomers; the numbers in parentheses indicate the number of isomers found in
each case (these are in agreement with the values obtained from the binary
icosahedral cluster atom counting rules in Ref.~\protect\onlinecite{TZKDS}).
The atom-exchange results are in qualitative agreement with the J-walking
results, but show increasingly large discrepancies as the temperature
decreases. The standard Metropolis algorithm was unable to reproduce the
peaks at all. The heat capacity peak areas as a function of cluster
composition are shown in Fig.~\protect\ref{Fig:Entropy}.
\label{Fig:Low_T_Cv}}
\end{figure}

\begin{figure}
\caption{Entropy changes associated with the low temperature permutational
isomerization transitions shown in Fig.~\protect\ref{Fig:Low_T_Cv} (circles).
These were obtained by numerically integrating the low temperature heat
capacity differences according to Eq.~\protect\ref{Eq:Delta_S} in the text.
Also shown for a qualitative comparison are $\Delta S/k_B = \ln N_{\rm Ar}$
(squares), where $N_{\rm Ar}$ is the number of low-lying Ar-core
icosahedral-like permutational isomers. The curves are symmetric about
Ar$_7$Kr$_6$.
\label{Fig:Entropy}}
\end{figure}

\begin{figure}
\caption{Solid-liquid transition heat capacity peak trends for thirteen-atom
Ar-Kr clusters. The top plot shows the transition peak temperatures as a
function of cluster composition, while the middle plot shows the transition
peak heights as a function of composition. The bottom plot depicts the peak
heights as a function of the peak temperature. The peak height error bars are
single standard deviations obtained from averaging the individual J-walker
runs; the standard deviations in the peak temperatures are smaller than the
symbol size. The peak temperature and height for each cluster also are listed
in Table~\protect\ref{Tbl:Cv_peaks}. The curves have been added merely as a
visual aid.
\label{Fig:Cv_Peaks}}
\end{figure}

\begin{figure}
\caption{Potential energy curves as functions of temperature for
Ar$_{13-n}$Kr$_n$ clusters ($0 \leq n \leq 13$). The curves were obtained
using J-walking.
\label{Fig:Caloric}}
\end{figure}

\begin{figure}
\caption{Quench results for the Ar-predominate clusters. These were obtained
by periodically quenching cluster configurations by steepest descent every
1000 passes during one of the J-walker runs, giving 1000 quenched
configurations for each temperature. The dotted vertical lines in each plot
indicate the solid-liquid heat capacity peak temperature, the dashed line in
the Ar$_{10}$Kr$_3$ plot the mixing-anomaly heat capacity peak temperature.
The dotted curve in the Ar$_{13}$ plot is the sum of the $-41.4720$,
$-41.4446$ and $-41.3944$ curves for the truncated icosahedral-like isomers
having one of the atoms displaced onto an icosahedral face. The curve is
similar to the corresponding curve shown in the Ar$_{12}$Kr plot representing
the sum of the topologically similar permutational isomers (energies ranging
from $-42.8345$ to $-42.5546$).
\label{Fig:Ar_Quench}}
\end{figure}

\begin{figure}
\caption{J-walker quench results for the Kr-predominate clusters. The
Ar$_3$Kr$_{10}$ curves are similar to the Ar$_{11}$Kr$_2$ curves at low
temperatures, reflecting their complementary permutations (Ar$_{11}$Kr$_2$
has three unique ways to position its two exterior Kr atoms, while
Ar$_3$Kr$_{10}$ has three unique ways to position its two exterior Ar atoms).
Likewise, the Ar$_4$Kr$_{9}$ and Ar$_{10}$Kr$_3$ curves are similar at low
temperatures, reflecting their five complementary permutations.
\label{Fig:Kr_Quench}}
\end{figure}

\begin{table}
\caption{Lennard-Jones parameters used in the calculations. The Ar-Ar and
Kr-Kr parameters were obtained from Ref.~\protect\onlinecite{LDW}. The Ar-Kr
parameters were obtained from the usual Lorentz-Berthelot mixing rules, with
$\epsilon_{\mbox{\scriptsize Ar-Kr}} = (\epsilon_{\mbox{\scriptsize
Ar-Ar}}\,\epsilon_{\mbox{\scriptsize Kr-Kr}})^{1/2}$ and
$\sigma_{\mbox{\scriptsize Ar-Kr}} = \frac{1}{2}(\sigma_{\mbox{\scriptsize
Ar-Ar}} + \sigma_{\mbox{\scriptsize Kr-Kr}}).$}
\begin{tabular}{lrrr}
\multicolumn{1}{c}{Parameter} & \multicolumn{1}{c}{Ar-Ar} &
\multicolumn{1}{c}{Ar-Kr} &
\multicolumn{1}{c}{Kr-Kr} \\
\hline
$\epsilon$/K & 119.4 & 139.9 & 164.0 \\
$\sigma/$\mbox{\AA} & 3.405 & 3.616 & 3.827 \\
\end{tabular}
\label{Tbl:LJ-params}
\end{table}

\widetext
\begin{table}
\caption{Potential energies for the thirteen lowest-energy equilibrium
configurations shown in Fig.~\protect\ref{Fig:Minima}. These values were
obtained from quench studies of the configurations stored in the J-walker
distribution files. The energies are expressed in units of
$-\epsilon_{\mbox{\scriptsize Ar-Ar}}$.
\label{Tbl:Minima}}
\begin{tabular}{*{8}{r}}
\multicolumn{1}{c}{Isomer} &
\multicolumn{1}{c}{Ar$_{13}$} &
\multicolumn{1}{c}{Ar$_{12}$Kr} &
\multicolumn{1}{c}{Ar$_{11}$Kr$_{2}$} &
\multicolumn{1}{c}{Ar$_{10}$Kr$_{3}$} &
\multicolumn{1}{c}{Ar$_{9}$Kr$_{4}$} &
\multicolumn{1}{c}{Ar$_{8}$Kr$_{5}$} &
\multicolumn{1}{c}{Ar$_{7}$Kr$_{6}$} \\
\hline
 0 & 44.32680 & 45.63591 & 46.96682 & 48.33095 & 49.67668 & 51.00709 & 52.36758 \\
 1 & 41.47198 & 42.83445 & 46.93058 & 48.28062 & 49.62872 & 50.94539 & 52.32164 \\
 2 & 41.44460 & 42.82518 & 46.91792 & 48.24762 & 49.58362 & 50.91443 & 52.30801 \\
 3 & 41.39440 & 42.80750 & 44.22434 & 48.23388 & 49.58077 & 50.91220 & 52.26145 \\
 4 & 40.75851 & 42.74808 & 44.21728 & 48.20980 & 49.56613 & 50.90030 & 52.22974 \\
 5 & 40.72846 & 42.74772 & 44.19670 & 45.65087 & 49.54682 & 50.89607 & 52.20526 \\
 6 & 40.71041 & 42.73919 & 44.13587 & 45.56089 & 49.53680 & 50.86535 & 52.20152 \\
 7 & 40.67380 & 42.70903 & 44.12606 & 45.54767 & 49.53207 & 50.85777 & 52.19877 \\
 8 & 40.67017 & 42.70607 & 44.12033 & 45.52440 & 49.52140 & 50.85168 & 52.19756 \\
 9 & 40.61547 & 42.69850 & 44.10625 & 45.51355 & 49.51229 & 50.83533 & 52.18107 \\
10 & 40.60458 & 42.68338 & 44.09642 & 45.50997 & 46.97535 & 50.83188 & 52.17493 \\
11 & 40.54129 & 42.67932 & 44.09642 & 45.47972 & 46.90413 & 50.82146 & 52.16898 \\
12 & 40.43333 & 42.66306 & 44.07858 & 45.46893 & 46.88614 & 48.29016 & 52.16469 \\
\hline
\multicolumn{1}{c}{Isomer} &
\multicolumn{1}{c}{Kr$_{13}$} &
\multicolumn{1}{c}{ArKr$_{12}$} &
\multicolumn{1}{c}{Ar$_{2}$Kr$_{11}$} &
\multicolumn{1}{c}{Ar$_{3}$Kr$_{10}$} &
\multicolumn{1}{c}{Ar$_{4}$Kr$_{9}$} &
\multicolumn{1}{c}{Ar$_{5}$Kr$_{8}$} &
\multicolumn{1}{c}{Ar$_{6}$Kr$_{7}$} \\
\hline
 0 & 60.88438 & 60.20496 & 58.85534 & 57.54709 & 56.26291 & 54.96607 & 53.65239 \\
 1 & 56.96319 & 59.26313 & 57.68168 & 57.48759 & 56.22254 & 54.92236 & 53.59395 \\
 2 & 56.92558 & 56.18509 & 57.63023 & 57.47117 & 56.16076 & 54.87811 & 53.57622 \\
 3 & 56.85663 & 56.15784 & 57.62012 & 56.12257 & 56.14611 & 54.86568 & 53.56163 \\
 4 & 55.98322 & 56.13239 & 55.36065 & 56.08925 & 56.10145 & 54.84450 & 53.54690 \\
 5 & 55.94194 & 56.10745 & 55.34242 & 56.03676 & 54.55578 & 54.81735 & 53.53000 \\
 6 & 55.91714 & 56.10706 & 55.28127 & 56.02826 & 54.51999 & 54.80451 & 53.50389 \\
 7 & 55.86686 & 56.02651 & 55.01280 & 55.98787 & 54.48346 & 54.80224 & 53.49223 \\
 8 & 55.86188 & 55.57558 & 55.00157 & 54.24192 & 54.47720 & 54.78820 & 53.48683 \\
 9 & 55.78674 & 55.56758 & 54.98270 & 54.23330 & 54.45869 & 54.75605 & 53.46915 \\
10 & 55.77179 & 55.54524 & 54.98117 & 54.21973 & 54.43440 & 52.97815 & 53.45425 \\
11 & 55.68486 & 55.53571 & 54.96068 & 54.21746 & 54.42764 & 52.93491 & 53.44179 \\
12 & 55.53657 & 55.50822 & 54.90435 & 54.20269 & 54.42467 & 52.91667 & 51.42744 \\
\end{tabular}
\end{table}
\newpage

\widetext
\begin{table}
\caption{Heat capacity peak parameters for thirteen-atom Ar-Kr clusters.
These values were obtained by smoothing and interpolating the J-walking data
shown in Figs.~\protect\ref{Fig:Cv} and \protect\ref{Fig:Low_T_Cv}. The
uncertainty estimates are averages of single standard deviations of the
points near the peaks.
\label{Tbl:Cv_peaks}}
\begin{tabular}{ldddd}
  &
\multicolumn{2}{c}{Mixing Anomaly Peak} &
\multicolumn{2}{c}{Solid-liquid Transition Peak} \\
\cline{2-3}
\cline{4-5}
  &
\multicolumn{1}{c}{$T_{\rm peak}$ (K)} &
\multicolumn{1}{c}{$\langle C_V^*\rangle_{\rm peak}$} &
\multicolumn{1}{c}{$T_{\rm peak}$ (K)} &
\multicolumn{1}{c}{$\langle C_V^*\rangle_{\rm peak}$} \\
\hline
Ar$_{13}$       &                 &
                & 34.05 $\pm$0.06 & 118.5 $\pm$0.6 \\
Ar$_{12}$Kr     &                 &
                & 34.41 $\pm$0.04 & 120.6 $\pm$0.6 \\
Ar$_{11}$Kr$_2$ &                 &
                & 34.94 $\pm$0.10 & 122.6 $\pm$0.5 \\
Ar$_{10}$Kr$_3$ &  2.74 $\pm$0.08 &  38.4 $\pm$0.2
                & 35.47 $\pm$0.10 & 123.6 $\pm$0.7 \\
Ar$_9$Kr$_4$    &  3.41 $\pm$0.08 &  38.9 $\pm$0.1
                & 35.86 $\pm$0.03 & 124.1 $\pm$0.8 \\
Ar$_8$Kr$_5$    &  4.38 $\pm$0.10 &  38.9 $\pm$0.1
                & 36.44 $\pm$0.18 & 123.8 $\pm$1.2 \\
Ar$_7$Kr$_6$    &  3.30 $\pm$0.03 &  39.5 $\pm$0.1
                & 37.05 $\pm$0.04 & 122.5 $\pm$0.9 \\
Ar$_6$Kr$_7$    &  4.14 $\pm$0.10 &  38.9 $\pm$0.1
                & 37.63 $\pm$0.10 & 119.9 $\pm$0.7 \\
Ar$_5$Kr$_8$    &  3.82 $\pm$0.23 &  38.5 $\pm$0.2
                & 38.60 $\pm$0.06 & 115.8 $\pm$0.9 \\
Ar$_4$Kr$_9$    &  3.40 $\pm$0.18 &  38.0 $\pm$0.1
                & 39.87 $\pm$0.09 & 112.1 $\pm$0.7 \\
Ar$_3$Kr$_{10}$ &                 &
                & 41.30 $\pm$0.12 & 110.2 $\pm$0.6 \\
Ar$_2$Kr$_{11}$ &                 &
                & 43.17 $\pm$0.09 & 111.0 $\pm$0.6 \\
ArKr$_{12}$     & 18.71 $\pm$0.15 &  48.6 $\pm$0.3
                & 44.86 $\pm$0.06 & 113.6 $\pm$0.4 \\
Kr$_{13}$       &                 &
                & 46.77 $\pm$0.09 & 118.0 $\pm$0.6 \\
\end{tabular}
\end{table}


\begin{references}
\bibitem{computers} All calculations were performed on either a DEC Alpha
    3000/300 AXP Unix workstation equipped with 64 Mb of RAM, a DEC Alpha
    3000/300L AXP equipped with 32 Mb of RAM, a DEC Alpha 3000/400 AXP,
    equipped with 64 Mb of RAM, or a Zeos Pantera~90 DOS PC, equipped with
    32 Mb of RAM.  The Zeos computer was equipped with a 90 MHz Pentium CPU
    that had the infamous hardware bug in floating point division. Given
    the stochastic nature of the calculations reported here and the
    reported infrequency of incorrect division results, it is believed that
    this bug has had no discernible effects in the numerical results.
    Some of the PC runs were repeated on a DEC workstation as a check
    and were found to be consistent.
\bibitem{acceptance} Distributions were typically used until the jump
    acceptance dropped down to about 10\%, although the values were often
    greater for the wider distributions in the solid-liquid transition
    region and slightly lower for the narrower distributions in the low
    temperature solid region.
\bibitem{Sav_Gol} The Savitsky-Golay algorithm was used for the data
    smoothing, and 2D cubic splines were used for the interpolation. These
    routines are implemented in the MS-DOS graphics software package Axum,
    version 3.0.
\bibitem{Kr-note} Since the constraining radius was constant at
    $3\sigma_{\mbox{\scriptsize Kr-Kr}}$ for all runs, the law of
    corresponding states implies that the Kr$_{13}$ simulations are
    equivalent to Ar$_{13}$ simulations using a smaller constraining sphere.
    Similarly, the other values listed were obtained from simulations using
    different constraining radii, and so minor discrepancies in the peak
    parameters are not unexpected.
\bibitem{low-T-JW} At such low temperatures, the J-walking runs effectively
    became Metropolis walks since none of the attempted jumps were accepted.
    More intermediate temperature J-walker distributions would have been
    needed to fully characterize this region.
\end{references}
\end{document}